# The Consequences of Eliminating NP Solutions[*]

*Piotr Faliszewski* and *Lane A. Hemaspaandra*

Department of Computer Science

University of Rochester

Rochester, NY 14627 USA

June 1, 2006; revised June 19, 2006 and August 16, 2006

**Abstract**

Given a function based on the computation of an NP machine, can one in general eliminate some solutions? That is, can one in general decrease the ambiguity? This simple question remains, even after extensive study by many researchers over many years, mostly unanswered. However, complexity-theoretic consequences and enabling conditions are known. In this tutorial-style article we look at some of those, focusing on the most natural framings: reducing the number of solutions of NP functions, refining the solutions of NP functions, and subtracting from or otherwise shrinking #P functions. We will see how small advice strings are important here, but we also will see how increasing advice size to achieve robustness is central to the proof of a key ambiguity-reduction result for NP functions.

## 1 Introduction

> Seal up the mouth of outrage for a while,
> Till we can clear these ambiguities.
> —Shakespeare, *Romeo and Juliet*, Scene 5, Act 3.

In everyday life it is natural to value clarity, and in particular to value those cases when a problem has a single, crisp answer. For example, if the question is, "Who won this presidential election?", not many people would be happy with the answer "Maybe Al Gore did and maybe George W. Bush did" or with the answer "Both Al Gore and George W. Bush did." And if a baseball team is deciding who will have the lead-off spot in the batting order, "Al and George will" is far from an acceptable outcome.

[*]This article is based on an invited talk at the Eighth Workshop on Descriptional Complexity of Formal Systems. Supported in part by grant NSF-CCF-0426761. URLs: http://www.cs.rochester.edu/u/pfali, http://www.cs.rochester.edu/u/lane. Appears also as URCS TR-2006-898.

In theoretical computer science, issues of ambiguity and multiplicity are also quite central. For a regular language, we may wonder about the size of the smallest unambiguous finite automata that accepts it. For a context-free language, we may ask whether it has an unambiguous context-free grammar. For an NP language, we may ask whether it is in UP, the unambiguous version of NP. For a multivalued NP function, we may ask whether it has a single-valued *refinement*—a pruning that on each instance on which the function has many solutions thins it down to exactly one solution.

In this paper, we will survey the issue of whether multiplicities can be reduced—whether solutions can be eliminated—in the case of functions involving NP machines. In particular, we will try to gather together and put in context for the reader as many of the known results on this as this article's space and flow constraints allow and, in some selected cases, will try to convey the flavor of representative proofs.

We will study the two most natural domains in which solution reduction is studied for functions based on NP machines: (a) reducing and refining solutions of NPMV functions (multivalued NP functions) and (b) reducing #P functions. It would be nice if all such reductions could be achieved with neither cost nor consequence. However, this seems not to be the case. Rather, in some cases solution reduction can be done and in other cases solution reduction can be shown to imply unlikely complexity-theoretic consequences.

This paper is structured as follows. Section 2 looks at solution reduction and solution refinement for NPMV functions. We will see that reducing solutions by polynomial-time computable, polynomially-bounded amounts is easy, but that refining solutions from many to one (or even from two to one) is impossible unless the polynomial hierarchy collapses. In the proof of this latter result we will see that small advice strings play a central role but that accepting an increase in advice size—in order to gain robustness—is the critical step in the proof. Section 3 focuses on the closure properties of #P—the class of functions that reflect the number of accepting paths of NP machines—with respect to decreasing operations, e.g., proper decrement, proper subtraction, minimum, integer division, etc. We will see that for each of these cases, #P is not closed under that type of decrease unless an unlikely complexity class collapse occurs.

Globally, we will take our alphabet $\Sigma$ to be $\{0, 1\}$, $\mathbb{N}$ will denote $\{0, 1, 2, \ldots\}$, $\mathbb{N}^+$ will denote $\{1, 2, 3, \ldots\}$, and for $n, m \in \mathbb{N}$, $n \mathbin{\dot{-}} m$ denotes $\max(n - m, 0)$.

## 2 Reducing and Refining NPMV Functions

This section focuses on whether we can reduce or refine the solutions of multivalued NP functions. Let us start by defining the classes and notions that will be needed.

### 2.1 Definitions

Let $f$ be a (potentially) partial, (potentially) multivalued function. (In mathematics, one would call $f$ a special kind of relation, but in theoretical computer science the term "multivalued function" is the norm in this context.) So, on each input $x$, either $f$ is undefined or some strings from $\Sigma^*$ are viewed as outputs of $f$. Following the standard notational approach to multivalued functions (see [Sel94]), we will set $\textit{set-f}(x) = \{y \mid y \text{ is an output of } f(x)\}$. Note in particular that $\textit{set-f}(x) = \emptyset$ if and only if $f(x)$ is undefined. The "$\textit{set-}$" notation avoids having to treat "undefined" as a special case, since it converts everything to clean output sets.

The classes of multivalued and single-valued NP functions, NPMV and NPSV, were introduced in the seminal paper of Book, Long, and Selman [BLS84]. These functions are structured as follows. Given a nondeterministic Turing machine $N$ having a designated output tape whose running time (i.e., the maximum number of steps of any of its computation paths) is polynomially bounded in the length of its input, on a given input we view each path that rejects as having no output. We view each path that accepts as outputting (a single string, namely) whatever string is between the fixed left-end marker of the machine's output tape and the cell under the output tape head (but not including the content of either of those cells). We view the machine, overall, as computing a potentially partial, potentially multivalued function $f$, where the outputs of $f$ on input $x$ are precisely the outputs of $N$'s accepting paths.

NPMV is the class of all partial, multivalued functions computed by nondeterministic polynomial-time machines. NPSV is defined as the set of all NPMV functions $f$ such that $(\forall x \in \Sigma^*)[\|\textit{set-f}(x)\| \leq 1]$. $\text{NPMV}_{\text{total}}$ (respectively, $\text{NPSV}_{\text{total}}$) denotes all $f \in$ NPMV (respectively, NPSV) such that $(\forall x \in \Sigma^*)[\|\textit{set-f}(x)\| \geq 1]$. These classes have been studied for many years, though under differing notations ([BLS84], and see also the excellent survey [Sel94]). What in this paper we for clarity denote as $\text{NPMV}_{\text{total}}$ and $\text{NPSV}_{\text{total}}$ are usually referred to simply as $\text{NPMV}_{\text{t}}$ and $\text{NPSV}_{\text{t}}$ in the literature.

Note that it is completely legal for different paths of an NP machine (modeling an NPMV function) to output the same value. However, since we view $set\text{-}f(x)$ as a set rather than a multiset, a given element is simply a regular member of $set\text{-}f(x)$ regardless of the (nonzero!) number of paths on which it is output. Also, note that it is completely legal for a multivalued function to happen to sometimes or always have only one output value—by multivalued we just mean "allowed to have multiple values."

We will be concerned with eliminating solutions from NPMV functions.

**Definition 2.1** *Given partial, multivalued functions $f$ and $g$, we say that $g$ is a fair reduction of $f$ exactly if $(\forall x \in \Sigma^*)[set\text{-}g(x) \subseteq set\text{-}f(x)]$.*

Of course, every partial, multivalued function $f$ is a fair reduction of itself, and the always undefined function is a fair reduction of all partial, multivalued functions. The latter fact makes clear why fair reductions are not the "right" notion to study: We want to prune down the number of outputs of multivalued functions—but certainly not to the point of eliminating all solutions. Rather, our dream case is to prune down from multiple solutions to one solution.

To capture the type of reduction we truly wish for, the right notion is not that of fair reduction, but rather is the notion of refinement.

**Definition 2.2 ([Sel94])** *Given partial, multivalued functions $f$ and $g$, we say that $g$ is a refinement of $f$, denoted $g \subseteq_c f$, exactly if $g$ is a fair reduction of $f$ and $(\forall x \in \Sigma^*)[set\text{-}g(x) \neq \emptyset$ iff $set\text{-}f(x) \neq \emptyset]$.*

That is, a refinement removes zero or more values from the output set, but never removes so many as to cross from having some outputs to having no outputs. It is true that each $f$ will trivially be a refinement of itself, but theorems about refinement generally block that case via conditions that ensure that, unless $f$ is NPSV to begin with, solution eliminations will occur (the exact nature of which will vary from theorem statement to theorem statement).

## 2.2 Fairly Reducing NPMV Functions by Small Amounts Can Be Done for Free

It is rather unfortunate that the natural goal is to understand refinements rather than to understand fair reductions. The reason it is unfortunate is that eliminating small numbers of solutions turns out to be easy and consequence-free for the case of fair reductions.

**Theorem 2.3** NPMV *is closed under fair reduction via proper subtraction of polynomial-time computable (or even* $\mathrm{NPSV}_{\mathrm{total}}$*-computable[1]), polynomially value-bounded numbers of solutions. (That is, if* $f \in \mathrm{NPMV}$ *and* $g \colon \Sigma^* \to \mathbb{N}$ *is a total, polynomial-time computable (or even* $\mathrm{NPSV}_{\mathrm{total}}$*-computable) function such that for some polynomial* $r$ *it holds that* $(\forall x \in \Sigma^*)[g(x) \leq r(|x|)]$*, then there exists some function* $h \in \mathrm{NPMV}$ *such that* $h$ *is a fair reduction of* $f$ *and, for each* $x \in \Sigma^*$*, it holds that* $\|\textit{set-}h(x)\| = \|\textit{set-}f(x)\| \mathbin{\dot{-}} g(x)$*.)*

**Proof.** Let $N$ be an NPTM modeling $f$. Compute $g(x) \in \mathbb{N}$. Since $g(x) \in \mathrm{NPSV}_{\mathrm{total}}$, we do this nondeterministically, and on all paths that compute an output value (and note that all those will compute the same output value) do the following. Nondeterministically guess a $1 + g(x)$ tuple of distinct computation paths of $N$ on input $x$. If all $1 + g(x)$ paths are accepting paths and have pairwise distinct outputs, then on the current path output the lexicographically largest output. Otherwise, the current path has no output (rejects). It is easy to see that we have just given an NP machine whose outputs on each $x$ are exactly all the elements of $\textit{set-}f(x)$ (if any) that are not among the $g(x)$ lexicographically smallest elements of $\textit{set-}f(x)$. ❑

So, for example, NPMV is closed under fair reduction via proper decrement. We will see in Section 3 that #P lacks that closure unless $\mathrm{NP} \subseteq \mathrm{SPP}$, and for the case of refinement we will see in Section 2.3 that even refining from (at most) two solutions to (at most) one would collapse the polynomial hierarchy.

There is a class, SpanP [KST89], that focuses on the number of distinct outputs of NP machines. Note that Theorem 2.3, due to its focus on cardinality reduction, is close to being a theorem about SpanP being closed under proper subtraction of simple, small functions (we have not been able to yet find that theorem in the literature, but it certainly seems a natural theorem that thus might be already known), except Theorem 2.3 is in fact stronger, since Theorem 2.3 is not merely reducing the "span" of an NPMV function, but is even doing so in a way that respects solution names (that is, that employs a fair reduction).

Exactly due to this connection between fair reductions and SpanP, it is certainly true that if NPMV is closed under fair reduction via proper subtraction of polynomial-time computable functions (note that we have removed the limitation that they be small in value), then SpanP is closed under proper subtraction of polynomial-time computable functions. Ogiwara and Hemachandra [OH93] have shown that the latter closure is completely characterized by the complexity class collapse $\text{NP} = \text{C} \cdot \text{NP}$ ("C·" has its usual literature meaning, namely, application of the counting operator associated with PP) or, equivalently, $\text{NP} = \text{PH} = \text{C}_=\text{P} = \text{PP} = \text{CH}$. So, certainly "NPMV is closed under fair reduction via proper subtraction of polynomial-time computable functions" implies that collapse. However, we observe that it is not hard to see that $\text{NP} = \text{PH} = \text{C}_=\text{P} = \text{PP} = \text{CH}$ implies "NPMV is closed under fair reduction via proper subtraction of polynomial-time computable functions." To see this, the crucial thing to note is that in $\text{PP}^{\text{NP}}$ we can accept the set that (for a fixed machine $N$ modeling an NPMV function, call it $f$) answers the question (the input to the set is $\langle x, n, y \rangle$) "On input $x$ is it the case that $y \in \text{set-}f(x)$ and there are at least $n$ elements in $\text{set-}f(x)$ that are lexicographically less than $y$." But $\text{PP}^{\text{NP}} \subseteq \text{CH}$, and so the assumption $\text{NP} = \text{CH}$ makes it easy to see the desired implication (similarly to the proof of Theorem 2.3, we can kill off an appropriate collection of lexicographically smallest output values). So putting together the previous comments, NPMV is closed under fair reduction via proper subtraction of polynomial-time computable functions exactly if $\text{NP} = \text{C} \cdot \text{NP}$. It is also easy to see that this remains true even if the subtracted functions are allowed to be $\text{NPSV}_{\text{total}}$-computable: Our algorithm can simply start off by $\text{NPSV}_{\text{total}}$-computing the number of solutions to remove, and then each path that successfully computes that value proceeds using the above approach.

---
[1]Technically, $\mathrm{NPSV}_{\mathrm{total}}$ functions map from $\Sigma^*$ to a single-valued subset of $\Sigma^*$, but via a bit of type coercion and the standard bijection between $\mathbb{N}$ and $\Sigma^*$ we may view them in this particular setting as a way of computing a mapping from $\Sigma^*$ to $\mathbb{N}$.

**Theorem 2.4** *The following conditions are equivalent.*

1. *NPMV is closed under fair reduction via proper subtraction of polynomial-time computable functions.*

2. *NPMV is closed under fair reduction via proper subtraction of $\text{NPSV}_{\text{total}}$-computable functions.*

3. $\text{NP} = \text{C} \cdot \text{NP}$.

Equivalently, in light of [OH93] and the above discussion, NPMV is closed under fair reduction via proper subtraction of polynomial-time computable functions (or even $\text{NPSV}_{\text{total}}$-computable functions) exactly if SpanP is closed under proper subtraction of polynomial-time computable functions. [OH93] provides about a dozen other statements that are equivalent to each of these statements, i.e., that are also characterized by $\text{NP} = \text{C} \cdot \text{NP}$.

Readers interested in the class defined by the closure of SpanP under subtraction (not proper subtraction, but subtraction; so this class will have functions that can take on negative values) are referred to the interesting work of Mahajan, Thierauf, and Vinodchandran [MTV94], which provides for SpanP a "gap"-like analog that parallels the relationship of GapP to #P.

## 2.3 Refining NPMV Functions to Unique Solutions Collapses the Polynomial Hierarchy, As Do Many Other Refinement Hypotheses

In this section we survey the known cases in which refining NPMV functions collapses the polynomial hierarchy.

Let NP2V denote all $f \in \text{NPMV}$ satisfying $(\forall x \in \Sigma^*)[\|\text{set-}f(x)\| \leq 2]$ [HNOS96]. Among the standard classes we will employ are PH (the polynomial hierarchy: $\text{P} \cup \text{NP} \cup \text{NP}^{\text{NP}} \cup \ldots$) [MS72, Sto76], ZPP (zero-error probabilistic polynomial time) [Gil77], $S_2$ (the symmetric version of $\text{NP}^{\text{NP}}$) [Can96, RS98], and $S_2^{\text{NP} \cap \text{coNP}}$ (all sets computable via relativizing $S_2$ with sets from $\text{NP} \cap \text{coNP}$) (see [CCHO05]). For completeness we give the definitions of $S_2$ and $S_2^{\text{NP} \cap \text{coNP}}$.

**Definition 2.5**    1. *([Can96, RS98]) A language $L$ is in $S_2$ if there exists a polynomial-time computable 3-argument boolean predicate $P$ and a polynomial $p$ such that, for all $x \in \Sigma^*$,*

     *(a) $x \in L \iff (\exists y \in \Sigma^* : |y| = p(|x|))(\forall z \in \Sigma^* : |z| = p(|x|))[P(x, y, z) = 1]$, and*

     *(b) $x \notin L \iff (\exists z \in \Sigma^* : |z| = p(|x|))(\forall y \in \Sigma^* : |y| = p(|x|))[P(x, y, z) = 0]$.*

   2. *([CCHO05]) A language $L$ is in $S_2^{\text{NP} \cap \text{coNP}}$ if there exists a 3-argument boolean predicate $P \in \text{NP} \cap \text{coNP}$ and a polynomial $p$ such that, for all $x \in \Sigma^*$,*

     *(a) $x \in L \iff (\exists y \in \Sigma^* : |y| = p(|x|))(\forall z \in \Sigma^* : |z| = p(|x|))[P(x, y, z) = 1]$, and*

     *(b) $x \notin L \iff (\exists z \in \Sigma^* : |z| = p(|x|))(\forall y \in \Sigma^* : |y| = p(|x|))[P(x, y, z) = 0]$.*

However, for the purposes of this paper, the only thing important to remember is that

$$\text{NP} \cup \text{coNP} \subseteq S_2 \subseteq S_2^{\text{NP} \cap \text{coNP}} \subseteq \text{ZPP}^{\text{NP}} \subseteq \text{NP}^{\text{NP}} \cap \text{coNP}^{\text{NP}}.$$

Also, we will need to draw on the notion of Karp–Lipton advice classes, in particular $(\text{NP} \cap \text{coNP})/\text{poly}$.

**Definition 2.6 (instantiating [KL80] to the case of $(\text{NP} \cap \text{coNP})/\text{poly}$)**
$(\text{NP} \cap \text{coNP})/\text{poly}$ *denotes each set $L$ such that there exists a set $A \in \text{NP} \cap \text{coNP}$*

and a polynomially length-bounded function $f\colon \Sigma^* \to \Sigma^*$ such that, $(\forall x \in \Sigma^*)[x \in L \iff \langle x, f(0^{|x|})\rangle \in A]$.

Informally put, there is an "advice-interpreter" set in $\mathrm{NP} \cap \mathrm{coNP}$ (this will soon come back to haunt us!) that with the right advice (which is short and depends only on the length of $x$) accepts exactly $A$. $(\mathrm{NP} \cap \mathrm{coNP})/\mathrm{quadratic}$ is analogous to Definition 2.6 except with $f$ required to be quadratically length-bounded.

We will draw on the following recent result, which we state without proof.

**Theorem 2.7 ([CCHO05])** $\mathrm{NP} \subseteq (\mathrm{NP} \cap \mathrm{coNP})/\mathrm{poly} \implies \mathrm{S}_2^{\mathrm{NP} \cap \mathrm{coNP}} = \mathrm{PH}$.

Finally, we will draw on two more items. The first is the notion of NPSV-selectivity, and the second is a relatively easy lemma linking refinement to NPSV-selectivity.

**Definition 2.8 ([HNOS96])** *A set $L$ is said to be NPSV-selective if there is a function $f \in \mathrm{NPSV}$ such that*

*(a) $(\forall x, y \in \Sigma^*)[\text{set-}f(x, y) \subseteq \{x, y\}]$ and*

*(b) $(\forall x, y \in \Sigma^*)[\{x, y\} \cap L \neq \emptyset \implies (\text{set-}f(x, y) \neq \emptyset \land \text{set-}f(x, y) \subseteq A)]$.*

**Lemma 2.9 ([HNOS96])** *The following are equivalent.*

1. *All NPMV functions have NPSV refinements.*

2. *All NP2V functions have NPSV refinements.*

3. *All NP sets are NPSV-selective.*

**Proof.** $1 \implies 2$ is immediate. Regarding $2 \implies 3$, note that for each NP set $L$ there clearly is an NP2V function $f_L$ such that $(\forall x, y \in \Sigma^*)[\text{set-}f_L(x, y) = L \cap \{x, y\}]$. Note that if $f_L$ has an NPSV refinement, that refinement proves under Definition 2.8 that $L$ is NPSV-selective. So $2 \implies 3$. Regarding $3 \implies 1$, let $f$ be an NPMV function computed by NPTM $N$. Consider the set $A = \{\langle x, s\rangle \mid x \in \Sigma^* \text{ and } b \in \{0,1\}^* \text{ and there is some accepting path of } N \text{ on input } x \text{ whose nondeterministic guess sequence has } s \text{ as a prefix}\}$. Note that $A \in \mathrm{NP}$. But if $A$ is NPSV-selective, via NPSV function $g$, we can (deterministically) check whether $N(x)$ has no nondeterministic guesses and if so we are done and we output the corresponding output if any, and otherwise we run $g(\langle x, 0\rangle, \langle x, 1\rangle)$ and every accepting path of that checks whether the guess sequence it found is the full guess sequence for that path, and if so outputs the corresponding output if any, and otherwise continues the self-reduction process one more level (i.e., on a path that found that $\langle x, 0\rangle$ is output by $g(x)$, move on to simulating $g(\langle x, 00\rangle, \langle x, 01\rangle)$). We always, in calling $g$, order its arguments $g(\langle x, \alpha\rangle, \langle x, \beta\rangle)$ when $\alpha \leq_{lex} \beta$. Note that the just-described process creates an NPSV refinement of $f$, so $3 \implies 1$. ❑

We finally must state the key link between refinement and hierarchy collapse.

**Theorem 2.10 ([HNOS96])** NPSV-selective $\cap$ NP $\subseteq$ (NP $\cap$ coNP)/poly.

We can now show that unique solutions collapse the polynomial hierarchy. The following result is due to [HNOS96], except it is stated here with the stronger conclusion that Theorem 2.7 (of [CCHO05]) gave it.

**Theorem 2.11 ([HNOS96] in light of [CCHO05])** *If all* NPMV *functions (or even all* NP2V *functions) have* NPSV *refinements, then* $S_2^{NP \cap coNP} = PH$.

**Proof of Theorem 2.11.** If all NPMV functions have NPSV refinements (or even all NP2V functions do) then by Lemma 2.9 all NP sets are NPSV-selective. So by Theorem 2.10, NP $\subseteq$ (NP $\cap$ coNP)/poly. So by Theorem 2.7, $S_2^{NP \cap coNP}$PH. ❑

We are done, except for the proof of Theorem 2.10. We do not formally prove that, but rather we will give a high-level exposition of the proof. (For a formal proof, see [HNOS96].) Suppose we are given an NP set $L$ that is NPSV-selective via NPSV function $f$. Without the loss of generality, $f$ is symmetric (i.e., $(\forall x, y \in \Sigma^*)[f(x, y) = f(y, x)]$)—otherwise replace $f$ with $f'(a, b) = f(\min(a, b), \max(a, b))$, which is symmetric and which can easily be seen to be an NPSV-selector for $L$ since $f$ is an NPSV-selector for $L$. Consider for some arbitrary $n \in \mathbb{N}$ the set $L_n = L \cap \Sigma^n$. Imagine each element of $L_n$ being a node in a tournament such that, for $a, b \in L_n$, $a \neq b$, there is an edge from $a$ to $b$ if and only if $b \in \text{set-}f(a, b)$. By a standard divide-and-conquer argument, originally used by Ko [Ko83] in the related context of showing that P-selective $\subseteq$ P/poly, it is easy to see (the first step is to eliminate half the graph by choosing as part of $S_n$ some node that points to at least half the other nodes—by counting, some such node must exist) that there will be a subset, $S_n$, of $L_n$ of cardinality at most $\lfloor \log_2(\|L_n\| + 1) \rfloor$ such that $L_n = \{y \in \Sigma^n \mid (\exists a \in S_n)[y \in \text{set-}f(a, y)]\}$. Since $\|L_n\| \leq 2^n$ and each string in $S_n$ has $n$ bits, clearly $S_n$ can be coded using $O(n^2)$ bits.

It would be wonderful to declare victory now, via claiming that we have an (NP $\cap$ coNP)/quadratic attack that on length $n$ inputs uses $S_n$ as the advice and sees whether an input $x$, $|x| = n$, beats at least one element of $S_n$. This at first would seem to work perfectly, but in fact there is a subtle yet severe problem.

That problem can be seen as follows. Imagine the allegedly "NP ∩ coNP" set of (NP ∩ coNP)/quadratic that on input $\langle x, B \rangle$ takes the advice $B = \{a_1, a_2, \ldots, a_k\}$ (allegedly $B = S_{|x|}$) and runs $f(x, a_1)$ and on each path that accepts (and thus chooses $x$ or $a_1$) runs $f(x, a_2)$ and on each path that accepts... etc., etc. At the end of this sequence, assuming $B = S_{|x|}$, each path that found an accepting path for each of the $k$ applications of $f$ definitively knows that $x \in L$ (namely, if at least one of the $k$ applications of $f$ had $x$ as an output) or knows that $x \notin L$ (namely, if none of the $k$ applications yielded $x$ as the output). The problem is the innocent-looking "assuming $B = S_{|x|}$"! Everything we just described works perfectly if $B = S_{|x|}$, i.e., if we are given the correct advice. But the definition of (NP ∩ coNP)/poly requires the advice interpreter to be an NP ∩ coNP set, and since the alleged advice is part of that set's input, that means we must be "NP ∩ coNP"-like *even when given lies for advice.* Informally put, "NP∩coNP"-like means having a machine that on each input has at least one path that accepts or rejects, each accepting or rejecting path must be correct, and paths also are allowed to neither accept nor reject. (The technical term for such "NP∩coNP machines" is "strong" computation (see [Lon82,Sel78]).) However, our selector $f$ is NPSV, not necessarily NPSV$_{\text{total}}$! If $B = S_{|x|}$ then all the $a_i$'s are members of $L$, and so each $f(x, a_i)$ is defined since $a_i \in L \implies \textit{set-}f(x, a_i) \neq \emptyset$ (by the definition of NPSV-selectivity). But if $B$ is untrue advice and contains some $a_i$ that does not belong to $L$ and $x \notin L$, then $f(x, a_i)$ may be undefined and our computation is not "NP ∩ coNP"-like and the proof is in shambles.

The fix is to trade advice size for robustness. Instead of quadratic-sized advice, $S_{|x|}$, let us instead require the advice to be $S_{|x|}$ plus, for each $a_i \in S_{|x|}$, a proof that $a_i \in L$. Since $L \in$ NP, such proofs exist. Now, our "NP ∩ coNP"-like attack is home free. Given the true advice, it does the right thing, as described above. Given a giant lie—a bunch of $a_i$'s not all accompanied by valid membership proofs—it will detect that it is being lied to, and will reject. And, most interestingly, given a subtle lie—an advice set $B' = \{a_1, a_2, \ldots, a_k\}$ whose $a_i$'s are all length $|x|$ strings and that all are accompanied by valid membership proofs in $L$ but such that $B'$ does not happen to have the property (which $S_{|x|}$ crucially does have) that $L_n = \{y \in \Sigma^n \mid (\exists a \in B)[y \in \textit{set-}f(a, y)]\}$—we will fail to detect that we are being lied to, but nonetheless will act in an "NP ∩ coNP"-like fashion, since the fact that each $a_i$ belongs to $L$ ensures that each application of $f(x, a_i)$ has an output. In brief, by going from quadratic to polynomial advice (the polynomial depends on the certificate size of $L \in$ NP), we made our advice interpreter robust enough to weather lies.

This completes the proof sketch for Theorem 2.10.

We conclude this section by briefly listing results more recent than Theorem 2.11 that show additional cases where refining solutions collapses the polynomial hierarchy. To do so, we will use the notation that for each $A \subseteq \mathbb{N}^+$, $\text{NP}_A\text{V}$ denotes all NPMV functions $f$ such that $(\forall x \in \Sigma^*)[\|\text{set-}f(x)\| \in A \cup \{0\}]$ [HOW02]. For example, $\text{NPMV} = \text{NP}_{\mathbb{N}^+}\text{V}$, $\text{NPSV} = \text{NP}_{\{1\}}\text{V}$, and $\text{NP2V} = \text{NP}_{\{1,2\}}\text{V}$. So Theorem 2.11 says $\text{NP}_{\{1,2\}}\text{V} \subseteq_c \text{NP}_{\{1\}}\text{V} \implies \text{S}_2^{\text{NP} \cap \text{coNP}} = \text{PH}$. The following results give other hypotheses sufficient to collapse the hierarchy, but unfortunately they collapse it to a level substantially higher than $\text{S}_2^{\text{NP} \cap \text{coNP}}$. The reason is that no analog of Theorem 2.10 is known for these cases, and so the proofs use weaker techniques such as direct quantifier exchange.

**Theorem 2.12 ([NRRS98])** *Let $k \in \mathbb{N}^+$. If $\text{NP}_{\{1,2,\ldots,k+1\}}\text{V} \subseteq_c \text{NP}_{\{1,2,\ldots,k\}}\text{V}$, then $\text{NP}^{\text{NP}} = \text{PH}$.*

**Theorem 2.13 ([Ogi96])** *Let $0 < \gamma < 1$. If $\text{NP}_{\mathbb{N}^+}\text{V} \subseteq_c \text{NP}_{\{1,2,\ldots,\lfloor \max(1,n^\gamma) \rfloor\}}\text{V}$ (here $n$ is the length of the input), then $\text{NP}^{\text{NP}} = \text{PH}$.*

Even more recently, the following flexible but complex cases have been established (note that Theorem 2.15 implies Theorem 2.12).

**Theorem 2.14 ([HOW02])** *Let $A, B \subseteq \mathbb{N}^+$ be nonempty. Suppose there exist four integers $c > 0$, $d > 0$, $e \geq 0$, and $\delta \geq 0$ satisfying the following conditions:*

1. *$d \leq c \leq 2d$ and $\delta < 2d - c$,*

2. *$c, 2d + e \in A$,*

3. *$c - \delta \leq \min\{i \mid i \in B\} \leq c$, and*

4. *$2d - (2\delta + 1) \geq \max\{i \in B \mid i \leq 2d + e\}$.*

*Then $\text{NP}_A\text{V} \subseteq_c \text{NP}_B\text{V}$ implies $\text{NP}^{\text{NP}} = \text{PH}$.*

**Theorem 2.15 ([HOW02])** *Let $k \geq 2$ and $d$, $1 \leq d \leq k - 1$, be integers. Let $A, B \subseteq \mathbb{N}^+$ be such that $\binom{k-1}{k-d} \in A$, $\binom{k}{k-d} \in A$, and $\max\{i \mid i \in B \text{ and } i \leq \binom{k}{k-d}\} \leq \lceil \frac{k}{d} \rceil - 1$. Then $\text{NP}_A\text{V} \subseteq_c \text{NP}_B\text{V}$ implies $\text{NP}^{\text{NP}} = \text{PH}$.*

We refer the reader to [HOW02] for a wide variety of corollaries that follow from Theorems 2.14 and 2.15, for proofs of these results, and for a discussion of how an even more general "lowness" result can unify and extend further these claims.

This section focused on cases where refining solutions implies hierarchy collapses. For completeness, we mention that there is a relatively general result known showing that in many cases one *can* nontrivially refine functions. Theorem 2.16 is proven by tuple trickery reminiscent in flavor to that used in the proof of Theorem 2.3.

**Theorem 2.16 ([HOW02])** *Let $A \subseteq \mathbb{N}^+$ and $B \subseteq \mathbb{N}^+$ be finite sets such that $A = \{a_1, \ldots, a_m\}$ with $a_1 < a_2 < \cdots < a_m$. If $\|A\| = 0$ or $(\exists b_1, \ldots, b_m : 0 < b_1 < \cdots < b_m)[\{b_1, \ldots, b_m\} \subseteq B$ and $a_1 - b_1 \geq \cdots \geq a_m - b_m \geq 0]$, then $\mathrm{NP}_A \mathrm{V} \subseteq_c \mathrm{NP}_B \mathrm{V}$.*

For example, Theorem 2.16 yields $\mathrm{NP}_{\{10,20,100\}}\mathrm{V} \subseteq \mathrm{NP}_{\{5,12,92\}}\mathrm{V}$.

## 3 Reducing #P Functions

The goal of this section is to survey known results regarding reducing the number of solutions of #P functions. For example, given two #P functions $f$ and $g$, is the function $h(x) = f(x) \mathbin{\dot{-}} g(x)$ aways a #P function? Alternatively, can we show some unlikely complexity-theoretic consequence that would follow were #P to be closed under proper subtraction? We will see that the latter seems to be the case.

The class #P was defined by Valiant [Val79]. A function $f$ is in #P if there exists a nondeterministic Turing machine $N$ such that, for each string $x$, $f(x)$ equals the number of accepting paths of $N$ on input $x$. Some typical examples of #P functions include the function that given a boolean formula $\phi$ returns the number of satisfying assignments of $\phi$, and the function that given a graph and a positive integer $k$ returns the number of distinct $k$-colorings of that graph.

We say that *$f$ is a closure property* if there exists a positive integer $i$ such that $f$ is a function from $\mathbb{N}^i$ to $\mathbb{N}$ [OH93]. Within this paper we will mostly be interested in the cases $i = 2$ (e.g., proper subtraction, integer division, 2-ary minimum) and $i = 1$ (e.g., proper decrement, integer division by two). One may consider this framework to apply also to functions that take as arguments not a tuple of natural numbers but rather a tuple of strings, in such cases implicitly invoking the standard bijection between $\mathbb{N}$ and $\Sigma^*$. In fact, we assume that coercions in either direction between $\mathbb{N}$ and $\Sigma^*$ are done implicitly via such a bijection whenever from the context it is clear that such coercing is needed, even if we do not explicitly mention it. $f$ is said to be a P-*closure property* if $f$ is a closure property and $f$ is computable in polynomial time.

**Definition 3.1 ([OH93])** *Let $f \colon \mathbb{N}^i \to \mathbb{N}$ be a closure property. We will say that #P is*

closed under $f$—or, equivalently, will say that $f$ is a closure property of #P—if it holds that $(\forall g_1, g_2, \ldots, g_i \in \text{\#P})[f(g_1(x), g_2(x), \ldots, g_i(x)) \in \text{\#P}]$.

Note that #P has some very natural closure properties. For example, #P is closed under addition and multiplication [Reg85]. Let $f$ and $g$ be two #P functions and let $N_f$ and $N_g$ be two nondeterministic Turing machines via which $f$ and $g$ are defined. #P is closed under addition because we can construct a nondeterministic Turing machine $N_{f+g}$ that on input $x$ nondeterministically chooses one of $N_f(x)$ and $N_g(x)$ to simulate and then nondeterministically performs that simulation. Closure under multiplication follows from the fact that on input $x$ we can first simulate $N_f(x)$ and then on every accepting path we can simulate $N_g(x)$. However, in this paper we are interested in those closure properties that have the potential to *decrease* values. Such closure properties include proper subtraction, proper decrement, integer division, minimum, etc. Note that we have to use *proper* subtraction/decrement and *integer* division since the codomain of #P functions is $\mathbb{N}$.

The nature of reducing the number of solutions of #P functions is quite different from the nature of reducing the number of outputs of NPMV functions. For example, if $f$ were an NPSV refinement of some NPMV function $g$, it would be totally legal for $f$ to have, on some (or even all) inputs, more accepting paths than $g$, provided that for a given input each of the accepting paths would output the same value. The same comment applies to fair reductions of solutions. In particular, by Theorem 2.3 we know that NPMV is closed under fair reductions via proper decrement. However, as we will see, if #P is closed under proper decrement, then an unlikely complexity class collapse occurs.

The study of the complexity of closure properties of #P (and some other function classes that we will not discuss here—SpanP, MidP, and OptP) was initiated by Ogiwara and Hemachandra [OH93]. One of the most important contributions of their work was introducing the concept of a "hard" P-closure property for a given function class and showing that #P does indeed have such "hard" P-closure properties. Let $f\colon \mathbb{N}^i \to \mathbb{N}$ be a closure property. We will say that $f$ *is a #P-hard P-closure property* [OH93] if $f$ is a P-closure property and the following implication is true: If #P is closed under $f$ then #P is closed under all polynomial-time computable closure properties.

Interestingly enough, #P is closed under every polynomial-time computable closure property if and only if #P is closed under every closure property $f$ such that $f \in \text{\#P}$.[2] While at first this might seem surprising, it in fact will be established by Theorem 3.2,

---
[2] Note the difference between $f$ being a closure property of #P and $f \in \text{\#P}$ being a closure property.

which itself has a relatively simple proof.

Results regarding closure properties of #P functions involve the classes UP [Val76], SPP [OH93,FFK94], PP [Sim75,Gil77], $C_=P$ [Sim75,Wag86], and $\oplus P$ [PZ83,GP86]. For the sake of completeness, let us briefly recall their definitions, in the form that is typically employed in our setting.

A language $L$ belongs to the class UP if there exists a polynomial $q$ and a polynomial-time computable binary predicate $R$ such that, for each $x \in \Sigma^*$,

1. $x \in L \implies \|\{y \in \Sigma^* \mid |y| = q(|x|) \land R(x,y)\}\| = 1$, and

2. $x \notin L \implies \|\{y \in \Sigma^* \mid |y| = q(|x|) \land R(x,y)\}\| = 0$.

Typically, UP is defined directly via unambiguous nondeterministic polynomial-time Turing machines. That is, a language $L$ belongs to UP if there exists an NP Turing machine $N$ that on input $x$ has exactly one accepting path if $x \in L$ and that has no accepting paths if $x \notin L$. However, the former definition stresses the similarity between UP and SPP. SPP is a generalization of UP. A language $L$ belongs to SPP if there exist two polynomials, $q$ and $p$, and a polynomial-time computable binary predicate $R$ such that, for all $x \in \Sigma^*$,

1. $x \in L \implies \|\{y \in \Sigma^* \mid |y| = q(|x|) \land R(x,y)\}\| = 2^{p(|x|)} + 1$, and

2. $x \notin L \implies \|\{y \in \Sigma^* \mid |y| = q(|x|) \land R(x,y)\}\| = 2^{p(|x|)}$.

A language $L$ belongs to PP if there exists a polynomial $q$ and a polynomial-time computable binary predicate $R$ such that, for each $x \in \Sigma^*$,

$$x \in L \iff \|\{y \in \Sigma^* \mid |y| = q(|x|) \land R(x,y)\}\| \geq 2^{q(|x|)-1}.$$

Similarly, a language $L$ belongs to $C_=P$ if there exists a polynomial $q$ and a polynomial-time computable binary predicate $R$ such that, for each $x \in \Sigma^*$,

$$x \in L \implies \|\{y \in \Sigma^* \mid |y| = q(|x|) \land R(x,y)\}\| = 2^{q(|x|)-1}.$$

A language $L$ belongs to class $\oplus P$ if there exists a nondeterministic polynomial-time Turing machine $N$ such that, for each $x \in \Sigma^*$, $x \in L \iff N(x)$ has an odd number of accepting paths.

---

In the former case #P is closed under $f$ and in the latter $f$ is simply stated to be a #P function of type $\mathbb{N}^i \to \mathbb{N}$, for some $i$.

Naturally, there is a close connection between the above definitions and definitions directly employing nondeterministic Turing machines. However, the above definitions often are more convenient in the setting of closure properties.

The following theorem shows that some simple operations that decrease the number of solutions of #P functions are in fact "the least likely" to be closure properties of #P among closure properties that plausibly could be closure properties of #P. *One can think of this as a loose parallel to the framework of* NP-*completeness theory, which shows that certain* NP *sets—*SAT*,* CLIQUE*, etc.—are "the least likely"* NP *sets to belong to* P.

**Theorem 3.2 ([OH93])** *The following statements are equivalent.*

1. #P *is closed under proper subtraction.*

2. #P *is closed under integer division.*[3]

3. #P *is closed under every nonnegative polynomial-time computable function (i.e., #P has every polynomial-time computable closure property).*

4. #P *is closed under every #P-computable function (i.e., #P has every #P-computable closure property).*

5. UP = PP.

So proper subtraction and integer division are #P-hard P-closure properties. See [OH93] for additional #P-hard P-closure properties.

We will give a fairly detailed overview of the proof of Theorem 3.2 (see also [OH93, HO02]). Clearly, 4 implies 3, and 3 implies both 1 and 2. We will argue in some detail that 1 implies 5, and then we will give overviews of the proofs that $2 \implies 5$ and that $5 \implies 4$.

Let us show that if #P is closed under proper subtraction then UP = PP. Assume that #P is closed under proper subtraction. First we will show that coNP $\subseteq$ UP and then we will show that PP $\subseteq$ NP. Since PP is closed under complementation, the latter implies that PP $\subseteq$ coNP. Thus, we have PP $\subseteq$ coNP $\subseteq$ UP, and since UP $\subseteq$ PP holds without assumption, we have UP = PP.

Recall that we have assumed that #P is closed under proper subtraction. Let $L$ be some arbitrary coNP language. By the definition of coNP, there exists a nondeterministic polynomial-time Turing machine $N_f$ such that if $x \in L$ then $N_f(x)$ has no accepting paths, and if $x \notin L$ then $N_f(x)$ has at least one accepting path. (In other words, $\overline{L}$ is an NP language.) Let $f$ be the #P function that $N_f$ implicitly defines. Since #P is closed under proper subtraction, $g(x) = 1 \mathbin{\dot{-}} f(x)$ is a #P function, and so there exists a nondeterministic Turing machine $N_g$ that on input $x$ has exactly $g(x)$ accepting paths. However, it is easy to see that

$$g(x) = \begin{cases} 1 & \text{if } x \in L, \\ 0 & \text{if } x \notin L. \end{cases}$$

---
[3]Note that for integer division, to avoid the issue of dividing by zero one would have to slightly adjust the notion of being closed under an operation. We will return to this issue later.

Thus, $N_g$ is a nondeterministic Turing machine that establishes $L \in \text{UP}$. (We use the fact that UP can be defined via unambiguous nondeterministic polynomial-time machines.) So coNP $\subseteq$ UP. (We mention in passing an alternate attack. Since below we will prove PP $\subseteq$ UP under our hypothesis, instead of proving coNP $\subseteq$ UP as we just did it would suffice to prove the weaker statement UP = NP, and that can be briskly seen, under our hypothesis, as follows: Given any NP machine $N_f$ and its associated #P function $f$, note that if #P is closed under proper subtraction then $f \mathbin{\dot{-}} (f \mathbin{\dot{-}} 1)$ must belong to #P, but this shows that $L(N_f)$ belongs to UP.)

Now, still under the assumption that #P is closed under proper subtraction, let us show that PP $\subseteq$ NP. Let $L$ be a PP language defined via a polynomial-time binary predicate $R$ and a polynomial $q$ such that for all $n$, $q(n) \geq 1$ and, for each $x \in \Sigma^*$,

$$x \in L \iff \|\{y \in \Sigma^* \mid |y| = q(|x|) \wedge R(x,y)\}\| \geq 2^{q(|x|)-1}.$$

Naturally, there exists a nondeterministic polynomial-time Turing machine $N_f$ that on input $x$ has exactly $\|\{y \in \Sigma^* \mid |y| = q(|x|) \wedge R(x,y)\}\|$ accepting paths. For example, on input $x$, this machine can guess a string $y$ of length $q(|x|)$ and then will accept on the current path exactly if $R(x,y)$ holds.

Let $f$ be the #P function implicitly defined by $N_f$, and let us consider the function $g(x) = f(x) \mathbin{\dot{-}} (2^{q(|x|)-1} - 1)$. Since $2^{q(|x|)-1} - 1$ is clearly a #P function and we assumed that #P is closed under proper subtraction, $g$ is a #P function as well. Thus, there is a nondeterministic polynomial-time Turing machine $N_g$ that on input $x$ has exactly $g(x)$ accepting computation paths. However, it is easy to see that if $x \in L$ then $g(x) \geq 1$ (as then $f(x) \geq 2^{q(|x|)-1}$) and that if $x \notin L$ then $g(x) = 0$ (as then $f(x) < 2^{q(|x|)-1}$). Thus, $N_g$ establishes $L \in \text{NP}$. Since $L$ was chosen arbitrarily, we have PP $\subseteq$ NP. This concludes the proof that if #P is closed under proper subtraction then UP = PP.

We will briefly argue as to why 5 implies 4. We assume that UP = PP and we will show,

separately for each fixed $i \in \mathbb{N}^+$, that given a #P function $f$, $f\colon \mathbb{N}^i \to \mathbb{N}$, and a sequence of $i$ #P functions $g_1, g_2, \ldots, g_i$ it holds that $h(x) = f(g_1(x), g_2(x), \ldots, g_i(x))$ is a #P function. Fix an arbitrary $i \in \mathbb{N}^+$. The idea behind the proof is to somehow get a handle on the values $g_1(x), g_2(x), \ldots, g_i(x)$ and then to directly run $f$ on those values.

As an aside, fix a nice multi-arity pairing function $\langle \cdots \rangle$ (see, e.g., [HHT97]). This function will of course take as arguments tuples of strings (but in fact, it is legal for, as is the case in the definition of $L_j$ for example, some components to be natural numbers, in which case the standard bijection between $\mathbb{N}$ and $\Sigma^*$ will be implicitly applied) and will output a single string. Returning to our proof, how can we obtain the values $g_1(x), g_2(x), \ldots, g_i(x)$? First, we observe that for each $j$, $1 \leq j \leq i$, the language $L_j = \{\langle x, y \rangle \mid g_j(x) \geq y\}$ is in PP. PP is closed under truth-table reductions [FR96]. So—though this actually requires the closure of PP just under bounded-truth-table reductions since $i$ is fixed—the language

$$L = \{\langle x, y_1, y_2, \ldots, y_i \rangle \mid (\forall j \in \{1, 2, \ldots, i\})[\langle x, y_j \rangle \in L_j \wedge \langle x, y_j + 1 \rangle \notin L_j]\}$$

is in PP as well. (We can decide $L$ using $2i$ queries to languages $L_1, L_2, \ldots, L_i$—two queries to each $L_j$, $1 \leq j \leq i$—each of which can be translated into a query to some selected PP-complete language.) Clearly, we have that $\langle x, y_1, \ldots, y_i \rangle \in L$ if and only if for all $j$, $1 \leq j \leq i$, it holds that $g_j(x) = y_j$.

Now, since UP = PP, we have that $L \in$ UP and so there is an unambiguous polynomial-time nondeterministic Turing machine $N$ that accepts $L$. Given this machine, we are ready to show that #P is closed under $f$. It is enough to construct a nondeterministic Turing machine $M$ that on input $x$ on each of its computation paths guesses a sequence of $i$ strings $z_1, z_2, \ldots, z_i$ (each $z_j$, $1 \leq j \leq i$, of appropriately polynomially bounded length) and simulates $N$ on input $\langle x, z_1, \ldots, z_i \rangle$. By the nature of $L$ and the fact that $N$ is unambiguous, there is only one path of $M$ that reaches $N$'s acceptance, and on that path we have complete information about the values $g_1(x), g_2(x), \ldots, g_i(x)$; it remains to simply run the nondeterministic polynomial-time Turing machine via which $f$ is defined with these values as input. This concludes the proof of the implication 5 $\implies$ 4.

Finally, we will quickly argue that if #P is closed under integer division then UP = NP. Note that in case of integer division we cannot really use our definition of what it means to be closed under an operation. If $f$ and $g$ are two #P functions then $\left\lfloor \frac{f(x)}{g(x)} \right\rfloor$ is undefined when $g(x) = 0$. Thus, we say that #P is closed under integer division if for every two #P functions $f$ and $g$ such that $(\forall x \in \Sigma^*)[g(x) > 0]$ the function $h(x) = \left\lfloor \frac{f(x)}{g(x)} \right\rfloor$ belongs to #P.

Let $L$ be a PP language. The key point here is that if $L \in$ PP then one can (with a bit of easy argumentation, which we will leave as an exercise, regarding item 3 below) claim that there is a positive integer $k$ and a machine $M$ that witnesses the membership of $L$ in PP that has the following properties:

1. On input $x$, $M$ has exactly $2^{|x|^k}$ computation paths, each containing exactly $|x|^k$ binary nondeterministic choices.

2. If on input $x$ $M$ has at least $2^{|x|^k-1}$ accepting paths then $x \in L$.

3. On each input $x$, $M$ has at most $2^{|x|^k} - 1$ accepting paths.

Let $f$ be the #P function defined by $M$. If #P is closed under integer division then $h(x) = \left\lfloor \frac{f(x)}{2^{|x|^k-1}} \right\rfloor$ is a #P function itself. However, we have that $h(x) = 1$ if $x \in L$ and $h(x) = 0$ otherwise. (We needed item 3 above to make sure that $h(x)$ is never greater than 1.) Clearly, the nondeterministic polynomial-time machine that defines $h$ is a UP machine that accepts $L$. This concludes our proof sketch of Theorem 3.2.

Note that Theorem 3.2 shows that if properly subtracting a #P function from a #P function always yields a #P function, then #P is closed under every #P-computable function. Can the hypothesis here be weakened? What if we instead assume just that properly subtracting a nonnegative polynomial-time computable function from a #P function always yields a #P function? Or what if we instead assume just that properly subtracting a #P function from a nonnegative polynomial-time computable function always yields a #P function? Are these two weaker assumptions still powerful enough to cause #P to be closed under every #P-computable function? The answer is that they both are indeed strong enough to yield that. That is, it turns out that each one of them is just as demanding an assumption as the seemingly stronger assumption that properly subtracting a #P function from a #P function always yields a #P function: All three of these assumptions stand or fall based on the same complexity-class equality. Of the two extended claims one is already known, and we state it without proof (Theorem 3.3). The other claim is new to this paper, and we prove it below as Theorem 3.4.

**Theorem 3.3 ([OH93])** UP $=$ PP *if and only if for every #P function $f$ and every polynomial-time computable function $g$ it holds that $h(x) = f(x) \mathbin{\dot-} g(x)$ is a #P function.*

**Theorem 3.4** UP $=$ PP *if and only if for every polynomial-time computable function $f$ and every #P function $g$ it holds that $h(x) = f(x) \mathbin{\dot-} g(x)$ is a #P function.*

To prove Theorem 3.4, we will need the following two facts regarding UP, PP, and C$_=$P.

**Proposition 3.5** 1. *[OH93, p. 304] UP = C$_=$P if and only if UP = PP.*

2. *[Sim75] $L \in$ C$_=$P if and only if there exists a polynomial $q$ and a polynomial-time computable binary predicate $R$ such that, for each $x \in \Sigma^*$,*

   (a) $x \in L \implies \|\{y \in \Sigma^* \mid |y| = q(|x|) \wedge R(x,y)\}\| = 2^{q(|x|)-2}$, *and*
   
   (b) $x \notin L \implies \|\{y \in \Sigma^* \mid |y| = q(|x|) \wedge R(x,y)\}\| < 2^{q(|x|)-2}$.

**Proof of Theorem 3.4.** The "only if" direction follows immediately from Theorem 3.2. We now will prove the "if" direction by showing that UP = C$_=$P follows if one assumes that for every nonnegative polynomial-time computable function $f$ and every #P function $g$ it holds that $h(x) = f(x) \mathbin{\dot{-}} g(x)$ is a #P function. Proving this suffices, since by Proposition 3.5 we know that UP = C$_=$P is equivalent to UP = PP.

Let $L$ be an arbitrary C$_=$P language. Let $R$ be the polynomial-time predicate and $q$ be the polynomial whose existence is guaranteed by Proposition 3.5. Clearly there exists an NP machine $M$ that, on arbitrary input $x$, guesses a string $y$ from $\Sigma^{q(|x|)}$ and accepts exactly if $R(x,y)$ holds. And so there is a #P function $f$—namely the function defined by the number of accepting paths of $M$—such that

1. if $x \in L$ then $f(x) = 2^{q(|x|)-2}$, and

2. if $x \notin L$ then $f(x) < 2^{q(|x|)-2}$.

By the "if" direction's assumption that properly subtracting #P functions from nonnegative polynomial-time computable functions only yields #P functions, we have that $g(x) = 2^{q(|x|)-2} \mathbin{\dot{-}} f(x)$ is a #P function. Note that $g(x) = 0$ if $x \in L$ and $g(x) > 0$ if $x \notin L$. Since $g$ is a #P function, by a second application of the "if" direction's assumption we have that $h(x) = 1 \mathbin{\dot{-}} g(x)$ is a #P function. But by these constructions,

$$h(x) = \begin{cases} 1 & \text{if } x \in L, \\ 0 & \text{if } x \notin L. \end{cases}$$

So each NP machine that instantiates $h \in$ #P is in fact a UP machine that accepts $L$. ❑

Theorem 3.2 shows that there are #P-hard P-closure properties. It is interesting to see that these hard closure properties are in fact very simple solution-reducing operations. As mentioned earlier, #P-hard P-closure properties can be viewed as analogs of complete languages, e.g., as analogs of NP-complete languages. On the other hand the closure properties that #P actually has (e.g., addition and multiplication), can be viewed as analogs of languages in P. By Ladner's Theorem [Lad75] we know that if P $\neq$ NP then there are also intermediate NP languages: Languages that are in NP − P but that are not NP-complete. Interestingly, there seem to be analogs of those in the world of closure properties as well. Some candidates for intermediate closure properties are proper decrement, minimum, and integer division by two. However, much as in the case of the graph isomorphism problem (which is suspected, but not known, to be NP-intermediate), we do not have a proof that these P-closure properties are either #P-hard or feasible, but rather have some pieces of evidence that suggest that that may be the case.

It would be nice to be able to *prove* that Theorem 3.2 in fact says that every closure property that isn't obviously not a closure property of #P becomes a closure property of #P if proper subtraction is a closure property of #P. The following statement, considered hand-in-hand with Theorem 3.2, makes it clear that that is indeed the case.

**Theorem 3.6** *Let $f$ be a closure property. If $f \notin$ #P, then #P is not closed under $f$.*

**Proof.** Let $f$ be a closure property such that $f \notin$ #P. Since $f$ is a closure property, for some $i \geq 0$ (and in fact, $i = 0$ is impossible if $f \notin$ #P, so let us suppose $i \geq 1$) $f$ maps from $\mathbb{N}^i$ to $\mathbb{N}$. Fix any nice arity-$i$ pairing function, $\langle \cdots \rangle$. For each $1 \leq j \leq i$, let $g_j(\langle n_1, n_2, \ldots, n_i \rangle) = n_j$. Note that each $g_j$ is a #P function. However, suppose that #P is closed under $f$. Then $f(g_1(\langle n_1, n_2, \ldots, n_i \rangle), g_2(\langle n_1, n_2, \ldots, n_i \rangle), \ldots, g_i(\langle n_1, n_2, \ldots, n_i \rangle))$ belongs to #P. But $f(g_1(\langle n_1, n_2, \ldots, n_i \rangle), g_2(\langle n_1, n_2, \ldots, n_i \rangle), \ldots, g_i(\langle n_1, n_2, \ldots, n_i \rangle))$ clearly equals $f(n_1, n_2, \ldots, n_i)$, and so $f(n_1, n_2, \ldots, n_i) \in$ #P, a contradiction, so our supposition that #P is closed under $f$ must be wrong. ❏

As we saw in Theorem 3.2, whether #P is closed under its #P-hard P-closure properties is fully characterized in terms of complexity class collapses by UP = PP. In the case of proper decrement, integer division by two, and minimum we do not have such complete characterizations.

**Theorem 3.7**  1. *(Due to Torán, as noted in [OH93].) If #P is closed under proper decrement, then NP $\subseteq$ SPP.*

2. *([OH93]) If UP = NP, then #P is closed under proper decrement.*

3. *([OH93]) If #P is closed under integer division by two (i.e., under the function $f(n) = \lfloor \frac{n}{2} \rfloor$), then SPP = $\oplus$P.*

4. ([OH93]) If #P is closed under minimum then UP = NP and SPP = $C_=P$.

The proofs of these theorems are similar in spirit to the proof of Theorem 3.2. They are based on the fact that each of the closure properties that we are dealing with has some kind of a conditional behavior built in that, together with some nondeterministic Turing machine trickery, can be exploited to cause a given complexity class collapse. For example, in the case of integer division by two, we can use it (together with the fact that #P is closed under multiplication by two) to decrement a function by one, provided that the function's value is odd. (We integer-divide it by two and then multiply it by two.)

In some cases, a complexity class equality can easily seem to be equivalent to a closure question. Here are two easy examples of that behavior, both of whose proofs are immediate from the definitions.

**Proposition 3.8**    1. #P is closed under $\min(1, n)$ if and only if UP = NP.

2. #P is closed under $1 \mathbin{\dot{-}} n$ if and only if UP = coNP.

**Proof.** The proof of the first equivalence is immediate and we will omit it. However, for the sake of completeness, let us quickly prove the second equivalence. As part of the proof of Theorem 3.2, following [HO02], we have already observed that if #P is closed under $1 \mathbin{\dot{-}} n$ then coNP $\subseteq$ UP. Clearly, coNP $\subseteq$ UP implies UP = coNP. It remains to show that if UP = coNP then #P is closed under $1 \mathbin{\dot{-}} n$. Let $f$ be an arbitrary #P function and let $L_f$ be the language $\{x \mid f(x) > 0\}$. Naturally, $L_f$ belongs to NP and so its complement, $\overline{L_f} = \{x \mid f(x) = 0\}$, is in coNP. Since we assumed that UP = coNP we have $\overline{L_f} \in$ UP, and thus there is a UP machine $M_h$ that accepts $\overline{L_f}$. We can view $M_h$ as instantiating a #P function $h$ such that

$$h(x) = \begin{cases} 1 & \text{if } f(x) = 0, \\ 0 & \text{if } f(x) > 0. \end{cases}$$

So the theorem is proven, since clearly $h(x) = 1 \mathbin{\dot{-}} f(x)$.   ❑

As a quick example of how equivalences such as those of Proposition 3.8 can help us understand the relationships between closure properties, consider the following assertion: If #P is closed under $1 \mathbin{\dot{-}} n$ then #P is closed under $\min(1, n)$. Faced with that assertion, one might not (though see the next paragraph) immediately know whether it was true. However, in light of Proposition 3.8 one instantly knows that the assertion is true, since the assertion becomes just another way of expressing the obvious fact that UP = coNP implies UP = NP.

The observations of the previous paragraph should be contrasted with the somewhat different—and very general and attractive (at least when it happens to work)—approach of using equalities to link closure properties in a way that spans all function classes. For example, for all natural numbers $n$ it clearly holds that $\min(1,n) = 1 \mathbin{\dot{-}} (1 \mathbin{\dot{-}} n)$. From that equality we may in one fell swoop conclude that for *every* class $\mathcal{F}$ of functions mapping from $\Sigma^*$ to $\mathbb{N}$ it holds that if $\mathcal{F}$ is closed under $1 \mathbin{\dot{-}} n$ then $\mathcal{F}$ is also closed under $\min(1,n)$.

Since in light of Theorem 3.7 proper subtraction is unlikely to be a closure property of #P, it is natural to seek to work around that by making the question a bit more flexible, and some papers have sought to do so. Building on the ideas of [OH93], interesting work of Gupta [Gup95, Gup92] has defined and studied analogous notions for the class GapP and for a quotient-based class he introduced, and also has introduced the notion of seeking not to exactly compute a closure but rather to approximate it with high probability on each input. It turns out that in these cases the results one gets are, loosely put, analogous to the vanilla #P cases. Ogihara et al. [OTTW96] have also suggested a somewhat different way of looking at closure properties of #P. We have seen that #P is not closed under proper subtraction. However, in a certain sense it is "close" to being closed under proper subtraction: Given two #P functions $f$ and $g$ one can easily see that there exists a polynomial $p$ (for example, any polynomial that is at least one more than the maximum of the nondeterminism polynomials of two fixed machines modeling $f$ and $g$) such that $h(x) = 2^{p(|x|)} + (f(x) - g(x))$ is a #P function. Thus, if we were allowed to perform some amount of postcomputation, then we could retrieve the value $f(x) \mathbin{\dot{-}} g(x)$ from $h(x)$.

## 4 Final Comments

In this paper, we looked at the issue of elimination of solutions in a few differing contexts. Though the paper is primarily tutorial-like, some results are to the best of our knowledge new to this paper, in particular Theorem 3.4, Theorem 3.6, Proposition 3.8, and all of Section 2.2.

Readers interested in the original or alternate treatments of the work covered in parts of Section 2 are referred to the various original literature papers cited in that section, and also to [HO02, HT03]. Among the most interesting related open issues are whether the collapses to $\text{NP}^{\text{NP}}$ can be strengthened to collapses to $\text{S}_2^{\text{NP} \cap \text{coNP}}$. Readers interested in the original or alternate treatments of the work covered in parts of Section 3 are similarly referred to the various original literature papers cited in that section, and also to [HO93,

HO02]. Among the most interesting related open issues are whether complete "complexity class collapse" characterizations can be found for the intermediate closure properties of #P. Fertile ground for additional research on the complexity of eliminating solutions includes the recently defined models of cluster-computed functions [HHKW,HHK06] (and [HHK06] starts in that direction by looking at decrementation) and interval functions [HHKW].